\documentclass[superscriptaddress,11pt]{article}
\usepackage{authblk}
\usepackage{array}
\usepackage{type1cm}
\usepackage{lettrine}
\usepackage{graphicx}
\usepackage{geometry}
\usepackage{psfrag}
\usepackage{amsmath}
\usepackage{amsfonts}
\usepackage{appendix}
\usepackage[margin=10pt,font=footnotesize,labelfont=sc,format=hang,labelformat=parens]%
{caption}
\usepackage{subcaption}
\usepackage{accents}
\usepackage{amssymb}%
\providecommand{\U}[1]{\protect\rule{.1in}{.1in}}
\numberwithin{equation}{section}
\def\be{\begin{equation}}
\def\ee{\end{equation}}
\def\ba{\begin{eqnarray}}
\def\ea{\end{eqnarray}}
\def\bi{\begin{itemize}}
\def\ei{\end{itemize}}

\def\bra{\langle}
\def\ket{\rangle}

\begin{document}

\title{Area discreteness, Lorentz covariance and Hilbert space non-separability }

\author{\normalfont Madhavan Varadarajan}
\affil{Dept. of Physics \& Astrophysics, University of Delhi,\\Delhi, India}
\maketitle

\begin{abstract}
We show how quantum discreteness of spatial area is consistent with  a unitary implementation
of Lorentz boosts in an LQG type quantization of a diffeomorphism invariant reformulation of 
free scalar field theory on 2d flat spacetime known as Parameterized Field Theory (PFT).
This consistency is a result of Hilbert space non-separability which is a characteristic feature of
LQG representations. Our results suggest a possible interpretation of Hilbert space non-separability
in terms of observer perspectives.
\end{abstract}

\thispagestyle{empty}
\let\oldthefootnote\thefootnote\renewcommand{\thefootnote}{\fnsymbol{footnote}}
\footnotetext{Email: madhavanvaradarajan@yahoo.co.in}
\let\thefootnote\oldthefootnote

\section{Introduction \label{sec1}}
In Loop Quantum Gravity (LQG) the area of a spatial surface is a well defined quantum operator with a discrete spectrum,  the lowest  eigenvalue being  zero with  a lowest non-zero eigenvalue of the order of the Planck area
\cite{area-rs-al}.
There is an intuitive tension between this discreteness and  local Lorentz covariance, specifically  covariance under {\em boosts}. In \cite{r-s}, it was noted
that just as discreteness of angular momentum in quantum mechanics is in no contradiction with  rotational
covariance, there is no in-principle tension between boost invariance and area discreteness. The key point 
is the continuity of operator {\em expectation values} despite the discreteness of the operator {\em spectrum}.

Since  boosts are transformations of both space and {\em time}, any analysis of Lorentz covariance necessarily 
involves the dynamics of the theory under consideration. 
Despite considerable progress,  the construction of a physically viable quantum dynamics for canonical LQG remains a difficult open problem. 
Hence, while  the detailed analysis of Reference \cite{r-s} 
builds a strong case for the consistency of Lorentz covariance with Area discreteness,
it is useful to explore this issue in a  simpler model context which shares the key feature of general covariance
with gravity and which admits a quantization similar to that underlying LQG
within which  its quantum dynamics is well defined. 1+1 d Parameterized Field Theory (PFT) is a  model system with precisely these properties.


PFT is a diffeomorphism invariant reformulation of free scalar field theory on $n+1$ dimensional flat spacetime
in which the flat spacetime  intertial coordinates are treated as dynamical variables to be varied in the action.
Consequently, in the Hamiltonian formulation these variables are canonical variables which are subject to quantization. 
PFT on 1+1 d spacetime admits an LQG type quantization resulting in a quantum physics which we refer to as Polymer Parameterized Field Theory  (PPFT)\cite{ppft2}.
\footnote{LQG type quantizations of model systems are referred to as `polymer' quantizations because    such 
quantizations are characterised by  quantum excitations  along {\em one dimensional} graphs and are  hence `polymer like'.}
  In 2d PPFT, the light cone combinations $X^{\pm}= T\pm X$ of the inertial time and space coordinates $(T,X)$ obtain discrete spectra in integer multiples of  fixed
parameters $a_{\pm}$ of dimensions of length. In \cite{ppft2,ppft3}, it is shown that the resulting  physics can be interpreted as  lattice field theory on a 2d light cone lattice i.e. on  {\em discrete} spacetime in which the lattice spacing in the $\pm$ directions is $a_{\pm}$.

In this work we shall focus on the implementation of boosts in the quantum theory. Whereas the work in \cite{ppft2,ppft3} focused on the sector $a_+= a_-=a$ with $a$ fixed, here we trivially extend those considerations to 
admit a certain one parameter set of $a_+, a_-$  sectors. The spectrum of the quantum spatial area (which in 1+1 d is the same as length) in each of these sectors is discrete and identical across sectors.
We show that, despite its  spacetime discreteness, PPFT supports a unitary representations of boosts which 
preserves the area spectrum.
As we shall see, the reason that the discrete spacetimes of PPFT support a unitary representation of boosts can be traced to the non-separable nature of the polymer representations used. We  argue
that such non-separability can be viewed in terms of observer perspectives in a precisely defined sense.
It is our hope that such a viewpoint might be usefully employed in the context of discussions of local Lorentz invariance in LQG.

In what follows we shall assume familiarity with the work \cite{ppft2} and confine ourselves to
an account of the minor generalizations/modifications of this work relevant to our considerations here.
These are two in number, with the first being the  admission of multiple sectors as described in the previous
paragraph, and the second being the extension of our considerations in these works (of  Minkowski spacetime with spatial topology $S^1$), to the case of usual planar Minkowski spacetime 
(with non-compact spatial topology  of the real line).

In section \ref{sec2} we extend the classical considerations of \cite{ppft2} to the case of planar spacetime
topology. Section \ref{sec3} describes the quantum kinematics. The kinematic Hilbert space is spanned by charge
network states each of which are eigen states of the embedding operators 
${\hat X}^{\pm}$. This Hilbert space consists of  sectors in which these operators have discrete spectra.
Each sector is  labelled by the  values of  Barbero-Immirzi \cite{b-i} like parameters which determine these spectra.
 In section \ref{sec4} we define and restrict attention to a superselected sector of the kinematic Hilbert space
 which we refer to as the finest lattice sector.
 \footnote{This sector is the generalization of the finest lattice sector
 in the spatially compact case \cite{ppft2}.}
  We discuss the classical asymptotic conditions in the context of 
 quantum states in
this sector. 
 We also construct the finest lattice sector physical Hilbert space through group averaging of finest lattice sector kinematic states.
In section \ref{sec5}  we define and analyse
  the action of boosts on finest lattice states both at the kinematic and the physical Hilbert space level.
  Section \ref{sec6} is devoted to a discussion of our results.

\section{\label{sec2} PPFT on planar spacetime: Classical Hamiltonian description}

\subsection{Classical Kinematics: The phase space variables}
The classical phase space variables  are fields on an abstract  1d Cauchy slice
$\Sigma$.
In the context of planar spacetime, 
$\Sigma$ is  diffeomorphic to the real line and  coordinatized by $x\in (-\infty, \infty)$. While we only need to fix this
coordinate system asymptotically near left and right  spatial infinity of the spacetime, it is convenient to fix this 
coordinate system on the entire real line once and for all.
Similar to Reference \cite{ppft2}, the canonical variables are the `embedding' variables $X (x), T(x)$. These
variables define an embedding of the abstract slice $\Sigma$ into 2d Minkowski spacetime $(M=R^2, \eta_{ab})$ by mapping  $x\in \Sigma$ to 
$(X (x), T(x)) \in M$ where $X,T$ are inertial coordinates for the flat spacetime metric $\eta_{ab}$.
The canonically conjugate embedding momenta are denoted by $P_X (x) , P_T (x)$.  The scalar field and its conjugate momentum are denoted by 
$\phi (x),\pi (x)$. 

In 2d, scalar field solutions to the wave equation on flat spacetime can be separated into left and right moving parts
which propagate along left and right moving light rays.
Consequently, the Hamiltonian dynamics simplifies in terms of  `left moving' and `right moving' canonical variables 
which are defined as follows.  The embedding variables $(X,P_X, T, P_T)$ are transformed to   the 
canonicially conjugate pairs 
$(X^+, \Pi_+), (X^-, \Pi_-)$  with $X^{\pm}(x)= T(x)\pm X(x)$ and $\Pi_{\pm}(x) =  \frac{1}{2} (P_T(x) \pm P_X(x))$
The matter variables are transformed to  the `left moving' and `right moving' combinations 
$Y^{\pm}(x) = \pi (x)\pm \phi^{\prime}(x)$ where the superscript `$\prime$' refers to derivation with respect to the spatial coordinate $x$. 
It can be checked that the `$+$' and `$-$' variables Poisson commute with each other with the new matter variables
Poisson brackets:
\be
\{Y^{\pm}(x), Y^{\pm}(y)\} = \pm (\frac{d\;}{dx} \delta (x, y)- \frac{d\;}{dy} \delta (y, x))
\label{y,y}
\ee

\subsection{Classical dynamics}
The classical dynamics is generated by the pair of constraints 
\begin{equation}\label{eq:2}
{{C_{\pm}}}(x)\ =\ [\ \Pi_{\pm}(x)X^{\pm'}(x)\ \pm\
\frac{1}{4} (Y^{\pm}(x))^2     \ ].
\end{equation}
These constraints are of density weight two. In 1 spatial dimension their transformation properties under
coordinate transformations are identical to those of  
spatial covector fields. Integrating them against multipliers $N^{\pm}$, which can therefore be thought of as 
spatial vector fields, one finds that the integrated `+' (respectively `-')  constraint generates 
spatial diffeomorphisms on the 
`+' (respectively `-') variables while keeping the `-' (respectively `+')  variables untouched. Thus, PFT dynamics can be
thought of as the action of two {\em independent} spatial diffeomorphisms $\Phi^+, \Phi^-$ on the 
`+' and `-' sectors of the phase space.

\subsection{Asymptotic conditions}
Since the spatial slice is non-compact, we need to specify the asymptotic behavior of the phase space variables
as $x\rightarrow \pm \infty$. For our purposes the following specification is adequate:
\ba
X^{\pm}(x)&=& \pm e^{\pm \lambda_R} x + \beta_{R,\pm} + O(1/x) , \;\;\;x\rightarrow \infty 
\label{xpmasymr}\\
X^{\pm}(x)&=& \pm e^{\pm \lambda_L} x + \beta_{L,\pm} + O(1/x) , \;\;\;x\rightarrow -\infty
\label{xpmasyml}
\ea
\ba
N^{\pm}(x)&=& \pm a_R x + b_{R,\pm} + O(1/x),  \;\;\;x\rightarrow \infty 
\label{npmasymr} \\
N^{\pm}(x)&=& \pm a_L x + b_{L,\pm} + O(1/x),  \;\;\;x\rightarrow -\infty
\label{npmasyml}
\ea
with the fields $\Pi_{\pm}(x), Y^{\pm}(x)$ required to be of compact support. Here $\lambda_{R}, \lambda_{L},\beta_{R,\pm},\beta_{L, \pm}$ and $a_R, a_L, b_{R,\pm},b_{L,\pm}$
are $x$-independent real parameters.

It can be checked that the equations of motion which are generated by the constraints smeared
with multipliers $N^{\pm}$ subject to the above asymptotic conditions preserve the asymptotic behavior of the 
phase space variables. We note here that in contrast to the case of gravity, the constraints smeared
with $N^{\pm}$ are themselves functionally differentiable without addition of any boundary terms despite the fact that the parameters $a_R, a_L, b_{R,\pm},b_{L,\pm}$ determine asymptotically non-trivial transformations
on $X^{\pm}$.

As remarked above, finite evolution is determined by the  pair of diffeomorphisms $\Phi_{+}, \Phi_-$
which, we note here, are generated by the 1d vector fields $N^{\pm}$ which are themselves subject to
the conditions (\ref{npmasymr}), (\ref{npmasyml}). In particular when the parameters $(a_R, a_L)$
are non-trivial the action of these diffeomorphisms  serve to  asymptotically boost  the spatial slices by 
independent left and right asymptotic boosts at, respectively, left and right infinity.  The difference between such 
diffeomorphisms and the boosts defined below in section \ref{cb} is that the former act on all phase space variables
and are gauge transformations whereas the latter act only on the embedding sector and are symmetries.

\subsection{Observables}
The spatial metric is 
\be
q_{xx}=-(X^-)^{\prime} (X^+)^{\prime},
\label{qxx}
\ee
 so that  the `Area' of a 1d  spatial `surface' $S$ is:
\be
A(S) =\int_S dx
\sqrt{-X^{-\prime} X^{+\prime}}
\label{area-c}
\ee
The area is a kinematic observable which does not commute with the constraints.

Let $f_{\pm}$ be compactly supported functions on the real line. Then the phase space functions
\be
O_{f_{\pm}}= \exp( {i\int_{-\infty}^{\infty} f_{\pm}(X^{\pm}(x)) Y^{\pm}(x)})
\label{ofc}
\ee
commute with the constraints and constitute a large set of Dirac observables. 
\footnote{This set may well be complete but a proof of this is beyond the scope of the paper.}

\subsection{\label{cb}Boosts}
Finite boosts correspond to the following finite canonical transformations of {\em only} the embedding variables with the matter variables left unchanged:
\ba
(X^+, P_+) &\rightarrow & (\lambda^{-1} X^{+}, \lambda P_{+}) \nonumber\\
(X^-, P_-) &\rightarrow & (\lambda X^{-}, \lambda^{-1} P_{-}) 
\label{boostc}
\ea
Here $\lambda$ is the $x$-independent  positive definite boost parameter. These transformations commute with the constraints. 

\section{\label{sec3}Quantum Kinematics: The charge network representation}
\subsection{The Embedding Sector}
The embedding sector Hilbert space is a tensor product of  `+' and `-' sectors. 
On the `$+$' sector the operator 
correspondents of functions on the `-' sector of phase space act trivially and vice versa.

The `+' embedding sector 
is spanned by an orthonormal  basis of  charge network states each of which is 
denoted by $|\gamma_{+}, \vec{ k^{+}} \ket$ where $\gamma_{+}$ is a graph i.e.
a set of edges which cover the real line  with  each edge $e_{}$ labelled by a `charge' $k^{+}_{e}$, 
the collection of  such charges for all the edges in the graph being denoted by $\vec{k^{+}}$.
Similar to the case of kinematic spin nets in LQG, two such states are orthogonal unless the 
edges (together with  their charge labels)  and the  vertices of the coarsest graphs underlying them 
coincide exactly (in which case the states are identical).
The $+$ embedding sector Hilbert space 
provides a representation of the Poisson bracket algebra between the classical `holonomy functions'
$\exp i(   \sum_{e} ( k_{e}^{+}\int_{e_{}}\Pi_+))$, and the embedding coordinate $X^{+}(x)$. 
In this representation the embedding momenta are polymerized so that  the holonomy functions 
are well defined operators but the embedding momenta themselves are not. The embedding coordinate
operator ${\hat X}^+(x)$ is well defined and the charge net states are eigen states of this operator.
In particular the action of the embedding coordinate operators ${\hat X}^{+}(x)$ on a charge network state
$|\gamma_{+}, \vec{k^{+}}\ket$
when the argument $x$ lies in the interior of an edge $e$ of $\gamma_+$ is:
\be
{\hat X}^{+}(x) |\gamma , \vec{k^{+}}\ket = \hbar k_e^{+} |\gamma , \vec{k^{+}}\ket 
\label{xplushat}
\ee
Identical results hold for $+\rightarrow -$. 
The  tensor product states $|\gamma_{+}, \vec{k^{+}}\ket\otimes |\gamma_{-}, \vec{k^{-}}\ket$ form a basis of the 
embedding sector Hilbert space and are referred to as embedding charge network states.
By going to a graph finer than $\gamma_+, \gamma_-$, each such state can be equally well denoted by 
$|\gamma, \vec{k^+}, \vec{k^-} \ket$ where each edge $e$  of the graph $\gamma$ is labelled by 
a {\em pair} of charges $(k^+_e, k^-_e)$ and the collection of these charges is denoted by $(\vec{k^+}, \vec{k^-})$.
Such a state is an eigen ket of both the ${\hat X}^+$ and ${\hat X}^-$ operators.
Similar to (\ref{xplushat}) the action of ${\hat X}^{\pm}(x)$ on the 
charge net $|\gamma, \vec{k^+}, \vec{k^-} \ket$ when $x$ is in the interior of  an edge $e$ of $\gamma$ is:
\be
{\hat X}^{\pm}(x) |\gamma , \vec{k^{+}},  \vec{k^{-}}\ket = \hbar k_e^{\pm} |\gamma , \vec{k^{+}},  \vec{k^{-}}\ket 
\label{xspect}
\ee
The charges are chosen to be integer valued multiples of a {\em fixed} dimensionful parameter $\frac{a}{\hbar}$ 
so that 
\be
\hbar k^{+}_e \in {\bf Z} \alpha a \;\;\;\;\;\;\;\;\;\;\;   \hbar k^{-}_e \in {\bf Z} \alpha^{- 1} a.
\label{xspect1}
\ee
Here the fixed parameter $a$ has dimensions of length and $\alpha$ is a real number. 

For fixed $\alpha, a$ the embedding holonomy functions $\exp i(   \sum_{e} ( k_{e}^{\pm}\int_{e_{}}\Pi_{\pm}))$
with $k^{\pm} _e$ subject to (\ref{xspect1}) form a complete set of functions. In this sense 
$\alpha, a$ are  Barbero- Immirzi like parameters which label unitarily inequivalent representations (unitary inequivalence follows from the distinct spectra for distinct $\alpha, a$ of the 
embedding operators).  The embedding Hilbert space is the linear sum of all these  $\alpha$-sectors
for all positive definite $\alpha$ and fixed $a$.

\subsection{\label{mqk}The Matter Sector}
The matter sector Hilbert space is also a tensor product of  `+' and `-' sectors. 
On the `+' sector the field $Y^{+}$ is polymerised and on the `-' sector the  $Y^-$ field is polymerised.
Thus taken together, neither $Y^+$ nor $Y^-$ 
exist as well defined operators.
The `+' sector provides 
a representation of matter holonomy functions
$\exp i(\sum_e l_e^+\int_eY_+ )$ on a basis of `+' matter charge nets, each such charge net denoted by 
$|\gamma_+, \vec{l^+}\ket$ in obvious notation. 
The `-' sector structure is identical
We fix the range of the matter charges $l^{\pm}$ to be such that:
\be
l^{\pm} \in {\bf Z} {\bf \epsilon}
\label{lepsilon}
\ee
for a fixed positive parameter  ${\bf \epsilon}$ of dimension $(ML)^{-\frac{1}{2}}$.
The holonomy functions  $\exp i(\sum_{e_{\pm}} l_{e_{\pm}}^{\pm}\int_eY_+ )$ with this restriction on $l_e^{\pm}$
are a complete set of functions on the matter phase space by virtue of the fact that the coordinate lengths of the edges of the graphs take values in the reals as opposed to the integers.

The algebraic structure on the matter holonomy operators is such that the 
$+$ and $-$ holonomies commute and for each of the $+$ and $-$ sectors, the holonomy  operators
satisfy a Weyl algebra by virtue of the Poisson brackets (\ref{y,y}) \cite{ppft2}. 
Accordingly, we denote the operator correspondent of  $\exp i(\sum_{e_{\pm}} l_{e_{\pm}}^{\pm}\int_eY_{\pm} )$
by ${\hat W}^{\pm}(\beta_{\pm}, {\vec l}^{\pm})$ where $\beta_{\pm}$ is the graph composed of the edges 
$\{e_{\pm} \}$ and ${\vec l}^{\pm}$  denotes the matter charge labels of these edges.

The  tensor product states $|\gamma_{+}, \vec{l^{+}}\ket\otimes  |\gamma_{-}, \vec{l^{-}}\ket$ form an orthonormal
 basis of the  matter Hilbert space and are referred to as matter charge net states.
By going to a fine enough graph any such state can be denoted, in notation similar to that for embedding states, 
as $|\gamma, \vec{l^+}, \vec{l^-} \ket$.
These states support the Weyl algebra of matter holonomy operators. The details of the operator
action will not concern us and the interested reader may consult \cite{ppft2}. Here it suffices to note that 
a $\pm$ matter holonomy operator augments the $\pm$ charges of the matter charge net it acts upon
by adding its matter charges to those of the charge net and  multiplies the resulting matter chargenet by 
an appropriate phase factor.
 
\subsection{The kinematic Hilbert space}
The tensor product of the matter and embedding Hilbert spaces yields the kinematic Hilbert space ${\cal H}_{kin}$ for PFT.
This Hilbert space is spanned by charge net states each of which is a tensor product of a matter charge net 
and an embedding charge net. By going to a fine enough graph underlying the matter and embedding charge nets 
we may denote such a tensor product state by 
$|\gamma, \vec{k_{\alpha}^+}, \vec{k_{\alpha}^-}, \vec{l^+}, \vec{l^-} \ket$ where, henceforth, we use the 
subscript $\alpha$ on the embedding charges to denote that they take values in the fixed $\alpha$-sector as
specified by (\ref{xspect1}).
Since the `+' and `-' sectors are independent, we also have the tensor product  decomposition:
\be
|\gamma, \vec{k_{\alpha}^+}, \vec{k_{\alpha}^-}, \vec{l^+}, \vec{l^-} \ket =  
|\gamma_+, \vec{k_{\alpha}^+}, \vec{l^+}\ket \otimes
|\gamma_-, \vec{k_{\alpha}^-}, \vec{l^-} \ket
\label{ptn}
\ee
  where 
$|\gamma_{\pm}, \vec{k_{\alpha}^{\pm}}, \vec{l^{\pm}}\ket$ is itself a 
product of a `$\pm$' embedding charge network and a `$\pm$' matter charge  network.
These states support the action of the embedding and matter holonomies, the former acting on the  underlying embedding charge network and  the  latter 
on the underlying matter charge network.

\subsection{The Area operator\label{areakin}}

Recall from (\ref{area-c}) that the classical `area' of a  `surface'  is 
$A(S) =\int_S dx\sqrt{-X^{-\prime} X^{+\prime}}$. We  restrict attention hereon to the case where $S$ is an 
{\em open}  interval of  the real line.
\footnote{ $S$ differs from its closure by its end points. These are sets of zero measure so it makes no difference
to the classical area whether we include them or not. However in the quantum theory due to the underlying 
discrete spectrum of ${\hat X}^{\pm}$ in each $\alpha$-sector, it is simplest to consider $S$ to be open and avoid potential complications arising from possible end point contributions.}

Using  the direction of increasing coordinate value, 
let the edge to the immediate left (right) of a vertex $v$ be denoted by $e_{v,L}$ ($e_{v,R}$) so that 
$e_{v,L} \cap e_{v,R}= v$. Then a straightforward  calculation similar to that in \cite{ppftham} shows that the only contributions to the area operator action
on a charge net arise from graph vertices and that this action evaluates to:
\be
{\hat A}(S)|\gamma, \vec{k_{\alpha}^+}, \vec{k_{\alpha}^-}, \vec{l^+}, \vec{l^-} \ket = \sum_{v\in S} \hbar \sqrt{|(k^{+}_{\alpha e_{v,R}}- k^{+}_{\alpha e_{v,L}}) (k^{-}_{\alpha e_{ v,L}}
- k^{-}_{\alpha e_{v,R}})|           }|\gamma, \vec{k_{\alpha}^+}, \vec{k_{\alpha}^-}, \vec{l^+}, \vec{l^-} \ket
\label{areahat}
\ee
Clearly, as a result of the condition (\ref{xspect1}) the area operator spectrum is {\em independent}
of the value of $\alpha$ and only depends on the parameter $a$. In this sense $a$ is the direct analog of the
Barbero-Immirzi parameter in LQG.

\subsection{Dirac Observables}
Consider the Dirac Observables  defined in (\ref{ofc}).  Since charge network states are eigen states of the 
embedding operators ${\hat X}^{\pm}$  it is straightforward to see that
\ba
{\hat O_{f_{\pm} } }|\gamma, \vec{k_{\alpha}^+}, \vec{k_{\alpha}^-}, \vec{l^+}, \vec{l^-} \ket &= &
{\hat W}^{\pm}(\gamma_{\pm}, {\vec f_{\pm}})
|\gamma, \vec{k_{\alpha}^+}, \vec{k_{\alpha}^-}, \vec{l^+}, \vec{l^-} \ket 
\label{ofchat}
\ea
Here ${\hat W}^{\pm}(\gamma_{\pm}, {\vec f_{\pm}})$ denotes the  matter holonomy operator (see section
\ref{mqk} for this notation)
based on the 
graph $\gamma_{\pm}$ (see (\ref{ptn})),  with the  edge $e_{\pm}$ of $\gamma_{\pm}$ labelled by a matter charge
given by the evaluation of $f^{\pm}$ on the {\em embedding} charge label of $e_{\pm}$ i.e. by 
$f^{\pm}(\hbar k_{\alpha e_{\pm}})$. 

In order that  this action  be well defined  we choose $f^{\pm}$  to be  functions from the real line to the 
integers (modulo the dimensional constant  $\epsilon$, see (\ref{lepsilon})). 
The only such functions which are smooth are the constant functions. 
Hence we relax the property of smoothness and  admit $f^{\pm}$ which are  {\em piecewise continuous}.
For simplicity, we restrict attention, as in equation (\ref{ofc}), to such  functions which are compactly
supported.

It is straightforward to check that the action (\ref{ofchat}) is unitary and consistent with the imposition 
of the classical `reality conditions' of these observables under complex conjugation (i.e. $O^*_{f^{\pm}} = O_{-f^{\pm}}$)
as adjointness conditions on their operator correspondents.

\subsection{Gauge transformations generated by the constraints.\label{sg}}

Recall that the finite transformations generated by the  constraints $H_+, H_-$ correspond to a pair
of diffeomorphisms $\Phi_+, \Phi_-$. The quantum kinematics supports a unitary  representation of 
these diffeomorphisms by the unitary operators ${\hat U}_+(\Phi_+ ), {\hat U}_-(\Phi_- )$.
The operator ${\hat U}_+(\Phi_+ )$ acts on a `+' charge network state 
$|\gamma_+, \vec{k_{\alpha}^+}, \vec{l^+}\ket$  by moving the graph and its colored edges by the diffeomorphism
$\Phi_+$ while acting as identity on `-' charge network states, and a similar action holds  for $+\rightarrow -$.
Clearly this action does not change the $\alpha$-sector.
We denote this action by 
\be
{\hat U}_{\pm}(\Phi_{\pm})|\gamma_{\pm}, \vec{k_{\alpha}^{\pm}}, \vec{l^{\pm}}\ket 
=: |\gamma_{\pm, \Phi_{\pm}}, \vec{k^{\pm}_{{\alpha}\Phi_{\pm}}}, \vec{l^{\pm}_{\Phi_{\pm}}}\ket
\label{uphipm}
\ee
The action of finite gauge transformations on a  charge net state 
$|\gamma, \vec{k_{\alpha}^+}, \vec{k_{\alpha}^-}, \vec{l^+}, \vec{l^-} \ket$ can then be deduced from equation (\ref{ptn}):
\be
{\hat U}_{+}(\Phi_{+}) {\hat U}_{-}(\Phi_{-})|\gamma, \vec{k_{\alpha}^+}, \vec{k_{\alpha}^-}, \vec{l^+}, \vec{l^-} \ket
= |\gamma_{+, \Phi_{+}}, \vec{k^{+}_{{\alpha}\Phi_{+}}}, \vec{ l^{+}_{ \Phi_{+} }}\ket \otimes
|\gamma_{-, \Phi_{-}}, \vec{k^{-}_{{\alpha}\Phi_{-}}}, \vec{l^{-}_{\Phi_{-}}}\ket .
\label{uphipphim}
\ee
By going to a graph finer than   $\gamma_{+, \Phi_{+}},  \gamma_{-, \Phi_{-}}$ the right hand side can 
again be written as a chargenet labelled by a single graph with each edge  labelled by a set of 4 charges
namely  the `+' and `-' embedding and matter charge labels.

In order to reduce notational complexity,
we shall  find it useful, on occasions,  to use the following condensed notation.
Recall that a  charge network in the $\alpha$ sector is denoted as in (\ref{ptn}).
We condense this notation and denote the charge network on the left hand side of (\ref{ptn}) by 
$s_{\alpha}$, the $\pm$ charge networks on the left hand side of (\ref{ptn}) by $s_{\alpha}^{\pm}$.
$s_{\alpha}^{\pm}$ is   the tensor product of an embedding and a matter charge network
which we denote by $s^{\pm}_{\alpha,emb}, s^{\pm}_{m}$.
In obvious notation we denote the images of these  under the gauge transformation $(\Phi_+, \Phi_-)$ by 
$s_{\Phi \alpha}$,    $s_{\Phi_{\pm}\alpha}^{\pm}$,  
$s^{\pm}_{\Phi_{\pm}\alpha,emb}, s^{\pm}_{\Phi_{\pm},m}$.

In this notation equation (\ref{uphipphim}) takes the form:
 \be
 {\hat U}_{+}(\Phi_{+}) {\hat U}_{-}(\Phi_{-})|s_{\alpha}\ket = |s_{\Phi \alpha}\ket
 \label{sphi}
 \ee

It is straightforward to verify that as expected the Dirac Observables ${\hat O}_{f^{\pm}}$ commute
with finite gauge transformations:
\be
[{\hat O}_{f^{\pm}},{\hat U}_{+}(\Phi_{+}) {\hat U}_{-}(\Phi_{-})] =0
\label{of,phi}
\ee

\section{\label{sec4}Implications of the asymptotic conditions and the finest lattice sector } 
Hitherto we have not discussed the implications of the classical asymptotic conditions in quantum theory.
For our purposes it suffices to discuss these in the context of a superselected sector of the kinematic Hilbert 
space which we call the finest lattice sector. We define this sector of the kinematic Hilbert space in section 
\ref{finlat} 
and its counterpart in the physical Hilbert space in section \ref{finlatp}.
Henceforth we shall restrict our attention to these sectors. Section \ref{justif} discusses a technical point related to the mutual orthogonality of distinct $\alpha$- sectors.

\subsection{\label{finlat}The finest lattice sector of the kinematic Hilbert space}
For each fixed $\alpha$ sector of the kinematic Hilbert space  consider the  subspace spanned by 
chargenets $|\gamma, \vec{k_{\alpha}^+}, \vec{k_{\alpha}^-}, \vec{l^+}, \vec{l^-} \ket= 
|\gamma_+, \vec{k_{\alpha}^+}, \vec{l^+}\ket \otimes
|\gamma_-, \vec{k_{\alpha}^-}, \vec{l^-} \ket$ subject to the following restrictions. \\

(i) Let  the number of edges of the  coarsest graphs $\gamma^{coarse}_{\pm}$ underlying 
$|\gamma_{\pm}, \vec{k_{\alpha}^{\pm}}, \vec{l^{\pm}} \ket$ be countable. Let us number these edges
 by an index $I^{\pm}$ which spans ${\bf Z}$
such that for $J^{\pm}>I^{\pm}$, the edge $e^{\pm}_{J^{\pm}}\in \gamma^{\pm}$ lies to the right of 
$e^{\pm}_{I^{\pm}}$  as a segment of the real line coordinatized by $x$
and such that the edges numbered consecutively share a vertex. The set of these edges cover the real line.\\

(ii) The  embedding charges on the coarsest graph $\gamma^{coarse}_{\pm}$ underlying 
$|\gamma_{\pm}, \vec{k_{\alpha}^{\pm}}, \vec{l^{\pm}} \ket$
 satisfy:
 \ba
 \hbar k^{+}_{{\alpha}e^{\prime}_{+}}-  \hbar k^{+}_{{\alpha}e_{+}} &=  &\alpha a \nonumber\\
 \hbar k^{-}_{{\alpha}e^{\prime}_{-}}-  \hbar k^{-}_{{\alpha}e_{-}} &=  & -(\alpha )^{-1}a 
 \label{kk+1}
\ea
where  $e^{\prime}_{\pm},e_{\pm}$ are adjacent edges in $\gamma^{coarse}_{\pm}$ such that 
$e^{\prime}_{\pm}$
lies to the right of $e_{\pm}$.\\

(iii) The  matter chargenet labels are `coarser' than the embedding ones so that each pair of successive edges
 of the coarsest graph
$\gamma^{coarse}_{\pm}$ underlying $|\gamma_{\pm}, \vec{k_{\alpha}^{\pm}}, \vec{l^{\pm}} \ket$ is necessarily labelled
by distinct pairs of $\pm$ embedding charges but not necessarily distinct pairs of $\pm$ matter  charges.\\
\noindent{\bf Note}:
The notion of  `coarsest' graph employed here is slightly different from the usual one employed in LQG like
representations
in the following sense.  Note that by virtue of equation (\ref{kk+1}) 
$\gamma^{\pm}$ will have exactly one  edge $e^{\pm}$ labelled by vanishing $k_{e^{\pm}}^{\pm}$.
In the usual LQG conventions such an edge is {\em not} part of the edges of the coarsest graph underlying the
embedding charge network. Hence if such an edge is labelled by a non-vanishing matter charge the matter
charge network labels would be considered {\em finer} than the embedding charge labels.
In a slight departure from this convention we define the notion of coarsest graph as one in which
consecutive edge labels differ even if edge labels for one of these edges vanish. Hence,  this notion
of coarsest graph for the states subject to equation (\ref{kk+1}) only requires that the edge 
$e^{\pm}$ labelled by vanishing $k_{e^{\pm}}^{\pm}$ be labelled by a {\em single} matter charge $l^{\pm}$.\\

(iv) Denote the length of an edge $e$ as measured in the fixed coordinate $x$ be $L_x(e)$.
Then there exists some integer $N_0>0$ and positive real numbers $\lambda_R, \lambda_L >0$ such 
that 
\ba
L_x(e^+_{I^+})&= & \lambda_R a\;\;\;\;\;\;L_x(e^-_{I^-})= \lambda_R^{-1}a, \;\; \forall I^{\pm}>N_0
\label{le=ar}\\
L_x(e^+_{I^+})&= & \lambda_L a\;\;\;\;\;\;L_x(e^-_{I^-})= \lambda_L^{-1}a, \;\; \forall I^{\pm}< -N_0
\label{ler=al}
\ea
To see how this condition captures the asymptotic behavior of $X^{\pm}$  (see (\ref{xpmasymr}), (\ref{xpmasyml}))let
us focus on the `right end' of the charge networks (similar considerations apply to the `left end').
Let the right vertex of $e^{+}_{N_0}$ be located $x=x_0^+$. From (\ref{le=ar}) the right  vertex of 
$e^{+}_{N_0+n}$ is located at $x^{+}_n = x^{+}_0 + n\lambda_R a$.
From 
(\ref{kk+1}) 
it then follows that:
\be
\hbar k^{+}_{{\alpha}N_0+n} =    \frac{\alpha}{\lambda_R}  x^+_n  - \frac{\alpha}{\lambda_R}x^+_0 +
\hbar k^{+}_{{\alpha}N_0}
\ee
Similar considerations yield 
\be
\hbar k^{-}_{{\alpha}N_0+n} =   - \frac{\lambda_R}{\alpha}  x^-_n  + \frac{\lambda}{\alpha}x^+_0 +
\hbar k^{-}_{{\alpha}N_0}
\ee
These conditions maybe looked upon as implementations of the classical conditions (\ref{xpmasymr}) in quantum
theory.\\

(v) In order to simplify the group averaging  procedure (see section \ref{finlatp}) it suffices for our purposes
to further require that:
\be 
l^{\pm}_{e_{I^{\pm}}}=0, \;\;\; \forall I^{\pm}< -N_0 \;\;\;{\rm and}\;\;\;\forall I^{\pm}>N_0 
\ee
\\

The finest lattice sector is then obtained as the finite span of chargenets for all $\alpha$, subject to (i)-(v) above
together with all their images obtained under the unitary action of all finite gauge transformations
as defined in section \ref{sg} (note that the finite diffeomorphisms $\Phi_{\pm}$ are generated by the vector
fields $N^{\pm}$ subject to the asymptotic behavior (\ref{npmasymr}),(\ref{npmasyml}).

We note that these gauge transformations look like  a combination of asymptotic boosts and translations
at right and left spatial infinity, a key  difference between these boosts and translations and the ones corresponding to  those described in section \ref{cb} are that the former act on both embedding and matter degrees of freedom
by diffeomorphisms whereas the latter act only on the embedding variables by a point transformation.

{\bf Note}: 
It is immediate to check that the embedding holonomy operators which are dependent on the embedding momenta
do not preserve the finest lattice sector.   Note that classically $X^{\pm}, C_{\pm}, Y_{\pm}$ can be used to reconstruct 
$P_{\pm}$. Hence, as in Reference \cite{ppft2} we adopt the view that the
classical functions whose counterparts we treat as primary are the constraints $X^{\pm}, C_{\pm}, Y_{\pm}$ rather than $X^{\pm}, C_{\pm}, Y_{\pm}$.  Note that the (finite transformations generated by) $C_{\pm}$ do
preserve the finest lattice sector.

\subsection{\label{justif}A note on the inter-sector inner product}
As seen above $\alpha$ sectors are superselected with respect to action of 
the basic operators which capture the physical content of  $X^{\pm}, C_{\pm}, Y_{\pm}$ (see the Note above).
In defining the kinematic Hilbert space in section \ref{sec3} to be the sum over the $\alpha$ sector
kinematic Hilbert spaces we have defined  the inner product on the kinematic Hilbert space to be such that
these sectors are mutually orthogonal.  Anticipating our definition  of boost operators in section \ref{sec5}, it 
will turn out that this inner product is consistent with  the unitary property of these boost operators.
We note here that independent of this  justification, the validity of this mutual orthogonality can also 
be viewed as a logical consequence of a larger Hilbert space structure as follows.

Consider, similar to Reference \cite{polypft}, the charges in the definition of the embedding holonomies to 
take independent values in the reals (modulo the factor of $\hbar$) instead of their values being in 
correspondence with the  integers in each fixed $\alpha$ sector. With this choice, the Hilbert space
is spanned by charge networks with real charges and charge networks 
which have distinct embedding charges on the same edge are orthogonal.  One can then notice
that within this large Hilbert space, there are $\alpha$-sectors which are superselected in the sense
defined in the paragraph preceding this one. The inner product on the large Hilbert space then 
implies that states in different $\alpha$-sectors are orthogonal.

\subsection{\label{finlatp}Quantum dynamics}
Recall that the finite transformations generated by the constraints are the diffeomorphisms 
$\Phi_{\pm}$ which are unitarily represented as in section \ref{sg}. As noted in section \ref{sg} 
the action of these unitary operators  do not change the
value of $\alpha$. Hence within each $\alpha$ sector we may group average kinematic charge network states 
with respect to this unitary action
of these  finite gauge transformations to obtain physical states. 
The group average \cite{alm2t}
$\eta (|s_{\alpha}\ket)$ of any such state
$|s_{\alpha}\ket$ (see the notation defined at the end of section \ref{sg}) is:
\be
\eta_{[s_{\alpha}]} \sum_{\Phi_+, \Phi_- \in Diff_{[s_{\alpha}]} }
\bra s_{\Phi \alpha}|
\label{orbit}
\ee
Here $[s_{\alpha} ]$ is the set of all distinct charge net states which are gauge related to $|s_{\alpha}\ket$ and 
$Diff_{[s_{\alpha}]}$ is a set of gauge transformations such that for every $|s^{\prime}_{\alpha} \ket \in [s_{\alpha}]$, 
there is precisely one gauge transformation which maps $|s_{\alpha}\ket$ to  $|s^{\prime}_{\alpha}\ket$.
$\eta_{[s_{\alpha}]}$ is a  parameter to be fixed by the requirement that the kinematic unitarity of the 
Dirac observables
${\hat O}_{f^{\pm}}$ holds for the group averaging inner product on the physical Hilbert space \cite{alm2t}.

{\bf Remark}:Since for any finest lattice charge network, the matter $\pm$ matter charge networks are finer than the 
$\pm$ embedding charge networks, it is straightforward to see that $Diff_{[s_{\alpha}]}$ can also be thought of as a 
set of gauge transformations such that for every $|s^{\prime}_{\alpha ,embed} \ket \in [s_{\alpha ,embed}]$, 
there is precisely one gauge transformation which maps $|s_{\alpha , embed}\ket$ to  
$|s^{\prime}_{\alpha , embed}\ket$. 

From the Remark above, it follows that   (\ref{orbit}) can be rewritten as:
\be
\eta_{[s_{\alpha}]} \sum_{\Phi_+, \Phi_- \in Diff_{[s_{\alpha , embed}]} }
\bra s_{\Phi \alpha}|
\label{orbite}
\ee


Using the dual action of operators on group averaged states  (\cite{alm2t}) we have that 
\ba
{\hat O}_{f^{\pm}} \eta (|s_{\alpha}\ket) &=: &\eta_{[s_{\alpha}]} \sum_{\Phi_+, \Phi_- \in Diff_{[s_{\alpha, embed}]} }
\bra s_{\Phi \alpha}|{\hat O}^{\dagger}_{f^{\pm}}\\
&=& \eta_{[s_{\alpha}]} \sum_{\Phi_+, \Phi_- \in Diff_{[s_{\alpha, embed}]} }
\bra ({\hat O}_{f^{\pm}}s)_{\Phi \alpha}| \\
&:=& \frac{\eta_{[s_{\alpha}]} }{\eta_{[{\hat O}_{f^{\pm}}s_{\alpha}]}}  \eta (|{\hat O}_{f^{\pm}}s_{\alpha}\ket)
\label{ocommute}
\ea
where in the second line $\bra ({\hat O}_{f^{\pm}}s)_{\Phi \alpha}|$ denotes the `bra' corresponding to the `ket'
$| ({\hat O}_{f^{\pm}}s)_{\Phi \alpha}\ket$ and in third line $[{\hat O}_{f^{\pm}}s_{\alpha}]$ refers to 
the gauge equivalence class of $| ({\hat O}_{f^{\pm}}s)_{\alpha}\ket$ . Further,  in the second line
we have used the fact that ${\hat O}_{f^{\pm}}$  is gauge invariant and in the third line that  its action on a charge network $|s_{\alpha}\ket$ 
does not change the embedding labels so that $s_{\alpha, embed}$ remains unchanged. 
It follows that the 
 choice 
 \be
 {\eta_{[s_{\alpha}]} }= {\eta_{[{\hat O}_{f^{\pm}}s_{\alpha}]}}
 \label{etas=etaos}
 \ee
 ensures that the action of the Dirac Observables ${\hat O}_{f^{\pm}}$ commutes with that
of the group averaging map  as required \cite{alm2t}. 

Note that by a suitable choice of $f^{\pm}$, in view of condition (v) of section \ref{finlat}, 
we can change the matter charges on any charge network to any prescribed  values. It then follows 
from (\ref{etas=etaos}) in conjunction with the Remark above that 
\be
{\eta_{[s_{\alpha}]} }= {\eta_{[s_{\alpha, embed}]} }
\label{eta=etaembed}
\ee

Finally with this simplification in the choice of the group averaging coefficients $\eta_{[s_{\alpha}]}$, it can be explicitly verified that 
 the reality conditions on   ${ O}_{f^{\pm}}$ are imposed as adjointness relations on their operator correspondents with respect to
the group averaging inner product:
\be
(\eta (|{s_{\alpha}}\ket), \eta (|s^{\prime}_{\alpha}\ket ) ):= \eta (|{s^{\prime}_{\alpha}}\ket)[|s_{\alpha}\ket]
\ee
where $\eta (|{s_{\alpha}}\ket )[|s^{\prime}_{\alpha}\ket]$ is the action of the distribution $\eta (|{s_{\alpha}}\ket)$
on the state $|s^{\prime}_{\alpha}\ket$ \cite{alm2t}.

For our purposes in this work we shall find it convenient to use the following  notation for the
physical state obtained by the group averaging of a kinematical state 
$|s_{\alpha}\ket=|\gamma, \vec{k_{\alpha}^+}, \vec{k_{\alpha}^-}, \vec{l^+}, \vec{l^-} \ket$:
\be
{(\gamma, \vec{k_{\alpha}^+}, \vec{k_{\alpha}^-}, \vec{l^+}, \vec{l^-} |}
= {\eta_{[s_{\alpha, embed}]} }\sum_{\Phi_+, \Phi_- \in Diff_{[s_{\alpha}]}}
\bra s_{\Phi \alpha}|
\label{physnot}
\ee

The finite span of all physical states  obtained by the group averaging of finest lattice charge network states
for all $\alpha$
constitutes the finest lattice sector of the physical Hilbert space.

\section{\label{sec5}Unitary implementation of boosts and Observer perspectives}
\subsection{\label{sec5.1}Boosts as Unitary operators and boost invariance of area}
We define the operator action ${\hat U}(\lambda)$  of a boost with boost parameter $\lambda$
on a  finest lattice state $|\gamma, \vec{k_{\beta}^+}, \vec{k_{\beta}^-}, \vec{l^+}, \vec{l^-} \ket$ in the 
$\beta$- sector as follows.
The operator action replaces the embedding charge labels $k^+_{\beta e}, k^-_{\beta e}$ for every edge $e$
of $\gamma$ by $\lambda k^+_{\beta e}, \lambda^{-1} k^-_{\beta e}$ leaving the matter charge labels 
unchanged. It can be checked that the resulting state is a finest lattice state that  lives on the same graph 
$\gamma$ but now belongs to the $\lambda \beta$ sector so that in equation (\ref{kk+1}) the value of $\alpha$
is changed from $\beta$ to $\lambda \beta$. We denote this action by:
\be
{\hat U}(\lambda)|\gamma, \vec{k_{\beta}^+}, \vec{k_{\beta}^-}, \vec{l^+}, \vec{l^-} \ket
=|\gamma, \vec{k_{\lambda \beta}^+}, \vec{k_{\lambda \beta}^-}, \vec{l^+}, \vec{l^-} \ket.
\label{boostkin}
\ee
It can be checked that this action is unitary by virtue of the orthogonality of states in distinct $\alpha$-sectors
(see discussion in section \ref{justif}),  commutes with the finite transformations generated by the constraints,
as well as  that of the matter holonomies and  that on any 
charge net in the finest lattice sector  the following relation holds
\ba
{\hat U}(\lambda) {\hat X}^{+} (x) {\hat U}^{\dagger}(\lambda)  &= &\lambda^{-1} {\hat X}^{+} (x),
\nonumber\\
{\hat U}(\lambda) {\hat X}^{-} (x) {\hat U}^{\dagger}(\lambda)&= &\lambda {\hat X}^{-} (x)
\label{boostxhat}
\ea
Thus the boost operator as defined in (\ref{boostkin}) implements the classical properties of 
boosts described in section \ref{cb} wherein we have adopted the viewpoint expressed in the Note at the
end of section \ref{finlat}.

As already mentioned in section \ref{areakin}, the Area operator spectrum is independent of the value of 
$\alpha$ which characterises each $\alpha$-sector. It is straightforward to check that equations (\ref{boostxhat}) 
and (\ref{areahat}) imply 
that the stronger property of
{\em invariance} of the Area operator under the action of boosts holds on the kinematic Hilbert space i.e.:
\be
{\hat U}( \lambda ) {\hat A}(S) {\hat U}^{\dagger} (\lambda) = {\hat A}(S)
\label{ahatinv}
\ee
This is a reflection of the classical boost invariance of the Area (\ref{area-c}) which follows from that of spatial
metric (\ref{qxx}) and the action of boosts on $X^{\pm}(x)$ (\ref{boostc}).
As we shall see in the next section the boost invariance of the classical area follows from general
considerations.

Next, we consider the action of boosts on physical states. 
As noted above  a boost with parameter $\lambda$ maps a finest lattice kinematic state in the $\alpha$ sector $|s_{\alpha}\ket$
to one in the $\lambda \alpha$ sector which we denote  $|s_{\lambda \alpha}\ket$. It is immediate to see that the action on  any 
gauge transformation in $Diff_{[s_{\alpha , embed}]}$  on  $|s_{\lambda \alpha}\ket$ yields  a distinct charge net
which is gauge related to $|s_{\lambda \alpha}\ket$, and that similarly, any 
gauge transformation in $Diff_{[s_{\lambda \alpha , embed}]}$  on  $|s_{\alpha}\ket$ yields  a distinct charge net
which is gauge related to $|s_{ \alpha}\ket$. This implies that we can choose
$Diff_{[s_{\lambda \alpha , embed}]}$ to be the same as $Diff_{[s_{\alpha , embed}]}$.
Using this together with the fact that the boost operators commute with finite gauge transformations
we have that:
\be 
{\hat U}( \lambda )(\gamma, \vec{k_{\alpha}^+}, \vec{k_{\alpha}^-}, \vec{l^+}, \vec{l^-} |
= \frac{\eta_{[s_{\alpha embed}]} }{ \eta_{[s_{\lambda \alpha embed}] } }
(\gamma, \vec{k_{\lambda \alpha}^+}, \vec{k_{\lambda \alpha}^-}, \vec{l^+}, \vec{l^-} |
\label{boostphys}
\ee
In order that boost commute with the group averaging map we set 
${\eta_{[s_{\alpha embed}]} }={ \eta_{[s_{\lambda \alpha embed} ]} }$.
\footnote{It seems plausible that there is only a single gauge equivalent class $[s_{\alpha embed}]$ in 
a fixed $\alpha$ sector. If this were true we could simply set ${\eta_{[s_{\alpha embed}]} }=1$
for all $\alpha$.}
With this choice it can be verified that the boost operators provide a unitary representation of  (the commutative
algebra of) boosts.

\subsection{Boost transformations of Dirac Observables\label{secboostdo}}
From (\ref{boostkin}), (\ref{boostphys}) it follows that both on the kinematic and the physical Hilbert space,
we have that:
\be
{\hat U}(\lambda ) {\hat O}_{f^{\pm}} {\hat U}^{\dagger} (\lambda)= 
 {\hat O}_{f_{\lambda}^{\pm}} 
\label{ofboost}
\ee
where $f^{\pm}_{\lambda}(X^{\pm}):= f^{\pm} (\lambda^{\mp 1} X^{\pm})$.
This is exactly the action of boosts at the classical level on the functions $f^{\pm}$.
\\
{\bf Note}: To see more intuitively how  this transformation results in a boost it is instructive to analyse this
transformation by relaxing our conditions on $f_{\pm}$ and setting $f^{\pm}(X^{\pm})$ to be the standard Fourier mode functions $e^{i \eta_{+-} p^{\mp} X^{\pm}}$ where
$\eta_{+-}= -\frac{1}{2}$ is the $\pm$ component  of the Minkowski metric $\eta$ in the $X^{\pm}$ 
coordinates.
The relationship between solutions of the free scalar wave equation $\box \phi (X^+, X^-)$ and the canonical
data $Y^{\pm}(x)$ turns out to be \cite{karelpft,polypft}:
\be
Y^{\pm}(x) = (X^{\pm})^{\prime}\frac{\partial \phi}{\partial X^{\pm}}|_{X^{\pm}= X^{\pm}(x)}=: 
(X^{\pm})^{\prime}\phi_{,\pm} (X^{\pm}(x))
\ee
where we have use the `$,\pm$' subscript to denote partial differentiation of the solution with 
respect to $X^{\pm}$. 
The exponent in the expression for $O_{f^{\pm}}$  then evaluates to the Fourier mode coefficients of
 $\phi_{,\pm}$:
\be
a_{\pm} (p^{\pm}) = \int e^{i \eta_{+-} p^{\mp} X^{\pm}} \phi_{,\pm} dX^{\pm}
\ee
The action of the boost (\ref{ofboost}) on this exponent yields:
\be
a_{\lambda,\pm} (p^{\pm}) = \int e^{i \eta_{+-} p^{\mp} \lambda^{\mp 1}X^{\pm} } \phi_{,\pm} dX^{\pm}
= a_{\pm} (\lambda^{\pm 1}p^{\pm})
\ee
as expected.

\subsection{Lattice interpretation of states in the finest lattice sector}
As outlined in \cite{ppft2,ppft3}, charge network states in the finest lattice sector have an immediate interpretation in
terms of matter excitations on  a discrete spacetime  lattice. To see this note
that every such  charge network state 
$|\gamma, \vec{k_{\alpha}^+}, \vec{k_{\alpha}^-}, \vec{l^+}, \vec{l^-} \ket$ is an
eigen state for the embedding operators ${\hat X}^{\pm}(x)$.
For $x$ in the interior of any edge $e$, these eigenvalues are $\hbar k^{\pm}_{\alpha\; e}$.
We can associate these eigenvalues to points in Minkowski spacetime through the set of light cone
coordinates $(X^+, X^-)= (\hbar k^+_{\alpha\; e}, \hbar k^-_{\alpha\; e}), \forall e \in \gamma$.
Just as the phase space variables $(X^+ (x), X^-(x))$ for all $x$ define an embedded Cauchy slice in Minkowski
spacetime,   the set of these points may be interpreted as defining a discrete Cauchy slice. 
Since the matter charges are coarser than the embedding ones, each such point $(\hbar k^+_{\alpha\; e}, \hbar k^-_{\alpha\; e})$ can be assigned a matter charge pair $(l^+_e, l^-_e)$ so that the charge net state admits
the interpretation of a matter excitations living on a discrete Cauchy slice.

Similarly, we can consider  the union of  all the points described by each charge network summand 
in the expression for the physical state 
$(\gamma, \vec{k_{\lambda \alpha}^+}, \vec{k_{\lambda \alpha}^-}, \vec{l^+}, \vec{l^-} |$ (see (\ref{physnot})).
It can be checked that each such point $p$ in Minkowski spacetime is labelled by the same pair of matter charges irrespective of 
the discrete Cauchy slices which contain this point $p$ i.e. irrespective of which summands describe these Cauchy slices
which contain $p$. As discussed in \cite{ppft2} this consistency of matter charge assignation can be traced to the fact that the kinematic Hilbert space supports an anomaly free representation of the finite transformations generated by the constraints.
It is straightforward to see that the set of these embedding charge points constitute a light cone lattice
discretization of Minkowski spacetime with lattice spacing in the $\pm$ directions being $\alpha^{\pm 1}a$.
Thus the physical state corresponding to a group average of a charge network state in the $\alpha$-sector
can be interpreted as corresponding to matter excitations on a light cone lattice with 
lattice spacings determined by $\alpha$.

\subsection{Interpretation of the action of boosts}
From the discussion in the previous section, boosts map  matter excitations on one spacetime lattice to  matter excitations
on a boosted spacetime lattice. Similar to the case of Fock space wherein the action of a boost on a momentum
eigen state can either be interpreted as a new state with boosted particle momenta or equally well
as the {\em same} state as perceived by a  boosted observer, here too  we can interpret 
the state obtained by the action of a boost in (\ref{boostkin}), (\ref{boostphys}) as a distinct boosted state
or as the same state as perceived by a boosted observer.

From (\ref{ofboost}), (\ref{boostphys}) and the Note in section \ref{secboostdo}, it follows that 
the right hand side of (\ref{boostphys}) can be thought of as a boosted state which when probed by
the boosted observable (\ref{ofboost}) yields the same state as that obtained by first probing the 
unboosted state with the unboosted observable and then taking the boosted image of the result.
Alternatively, we can think of (\ref{boostphys}) as describing the unboosted state as seen by a
 boosted observer. In this case the original observable for this observer is described by
(\ref{ofboost}) and the descriptions  of the same physical measurements on the same physical state
from the perspectives of the original and boosted observer 
 are unitarily related.

The reason that the discrete spacetime lattices for finest lattice states support a representation
of boosts can be traced to the fact that for every $\alpha$ we have an orthogonal sector of the Hilbert space.
Hence the existence of a unitary representation of boosts  in the finest lattice sector is a consequence
of the {\em non-separability} of the physical Hilbert space. Since the action of a boost can be interpreted
as describing the same physics from the perspective of a boosted observer, one could
perhaps take a stronger view that a state in a fixed $\alpha$-sector defines both a physical state as well
as a fixed observer. More in detail, let us restrict attention to a state in, say, the $\alpha=1$ sector and interpret 
measurements of  the Dirac observables ${\hat O}_{f^{\pm}}$ in this  sector as those seen by the
associated observer.  The description of the same physical measurements  by  a boosted 
observer at boost parameter $\alpha$ is obtained through unitary transformations of observables and state to 
the  $\alpha$ sector. Hence the entire physical content of the state can be extracted by staying in any fixed
$\alpha$ sector. 

\section{\label{sec6}Discussion}
Despite an intuitive tension between boost invariance and discreteness of area (which is the same as length in the 2d model
anlysed in this work), we have seen in section \ref{areakin} that the area operator spectrum is both discrete and 
boost invariant. Indeed, as noted in  section \ref{sec5.1}, equation (\ref{ahatinv}), the area operator is  itself 
boost invariant. Though perhaps physically counterintuitive, this invariance is not mathematically surprising given that the  expression (\ref{area-c}) is boost invariant by 
inspection.  As may  be expected, there is  an underlying reason for this boost invariance arising from 
 the fact that boosts are isometries of the Minkowski
spacetime metric. To see this note that the spatial metric on an  embedded slice $\Sigma$ in flat spacetime
is just the pull back $\underleftarrow{\eta}$ of the flat spacetime metric $\eta$ to the slice. The action of an isometry ${\cal I}$
on the slice moves the slice to its image by the isometry. The spatial metric on the new slice is the pull back of
the spacetime metric $\eta$ to the new slice ${\cal I}$ . But since ${\cal I}$ is a diffeomorphism which preserves $\eta$
we have that 
\be
A(S)= \int_S \sqrt{\det\underleftarrow{\eta}} = \int_{{\cal I}(S)} \sqrt{\det\underleftarrow{{\cal I}_*\eta} }
= \int_{{\cal I}(S)} \sqrt{\det\underleftarrow{\eta}} = A({\cal I}(S))
\ee
In the context of local Lorentz transformations in general relativity, the same argument would apply to sufficiently small surfaces in a convex normal neighborhood of a point. In this sense the classical area of sufficiently 
small surfaces is a {\em locally Lorentz invariant quantity}.

In this context while the work \cite{r-s} shows that there is no in-principle  contradiction between area discreteness  and  local Lorentz covariance, the recovery of  classical local Lorentz {\em invariance} of sufficiently small areas from the underlying
spatial discreteness of LQG is a very involved problem whose elucidation requires a thorough understanding
of the quantum dynamics of LQG as well as its semiclassical states. Despite significant progress these are as yet open problems at the frontier of LQG research. 
The issue has several facets and we hope that the 
existence  of a  connection noted here, admittedly in a simple toy model context, between observer perspectives, Lorentz covariance of discrete spacetime structures and Hilbert space non-separability (which is ubiquitous in the context of LQG representations) might 
prove helpful.

The work here also serves to emphasize that Lorentz {\em covariance} does not necessarily imply Lorentz 
{\em invariance}. Indeed, if we adopt the `observer perspective' point of view and restrict our attention
to a single $\alpha$-sector, it is difficult to define what one would mean by Lorentz invariance. However, 
due to the regular lattice structure of quantum spacetime, it seems impossible to construct a state in the
fixed $\alpha$ Hilbert space which is Lorentz invariant in any physically compelling sense.
For example, 
we could define a Lorentz invariant state  in a fixed $\alpha$- sector as one for which the expectation value of any
${\hat O}_{f^{\pm}}$ operator is the same as that of its boosted image (\ref{ofboost}). It is then  straightforward to see that 
for any $f^{+}$ compactly supported away from the origin and whose support extends over a few lattice spacings in the $X^+$ direction, a boost which shrinks the support  
to less than a lattice spacing, effectively converts the operator to the identity operator on this fixed $\alpha$ sector. 
It is then straightforward to check  that while the expectation value for the operator ${\hat O}_{f^+}$ vanishes the expectation value for its boosted counterpart  does not, as this boosted counterpart acts as the identity operator. This is 
simply because the boosted function is forced to probe sub-lattice scales and such scales are absent in the lattice. 

Indeed, the only context we are aware of in which the existence of discrete spacetime structures 
is consistent with Lorentz {\em invariance} is that of Causal Sets \cite{jlr} wherein spacetime discreteness is allied 
with the key property of {\em randomness} \cite{licauset}.  In the context of a Hamiltonian formulation, it seems to us
that it is necessary to have a {\em stochastic} component to the dynamical law. Whether this can be incorporated 
into the treatment of the Hamiltonian constraint in LQG is an intriguing open question.

On an unrelated note, another open issue in LQG relates to a satisfactory development of an LQG kinematics
which adequately  incorporates the classical conditions of asymptotic flatness. In this context it is of interest
to improve our treatment of asymptotic boundary conditions in the simple toy model setting of this work.
We have only 
provided a schematic treatment as this sufficed for our purposes. It would be   of interest to carefully articulate
both the classical boundary conditions on the embedding momenta and the matter fields in such a way that 
they can be imposed satisfactorily in the quantum theory perhaps along the lines of Reference \cite{ksasymp}. 
Even with regard to the imposition of the conditions on $X^{\pm}$ as conditions on states, it would be desireable
to understand in quantitaive detail how  the states obtained as gauge transformations of the ones specified by condition (iv)
of section \ref{finlat}  satisfy the asymptotic boundary conditions. It would also be of interest to 
explicitly parameterize the gauge invariant information in the group averaged state (\ref{physnot})
\footnote{While we expect that each fixed $\alpha$-sector of the physical Hilbert space is separable,
an explicit paramaterization is needed in order to confirm this expectation.}
for  example in the language of sequences of embedding and matter charges with a view to a generalization
for the case of LQG.

\section*{Acknowledgements:} I am grateful to Seth Major for his comments on a preliminary draft of this work.

\end{document}